\documentclass[10pt,twocolumn,letterpaper]{article}
\usepackage{iccv}

\usepackage{times}
\usepackage{epsfig}

\usepackage[accsupp]{axessibility} 

\usepackage{graphicx}
\usepackage{amsmath}
\usepackage{amssymb}
\usepackage{booktabs}
\usepackage[table,xcdraw]{xcolor}         
\usepackage{multirow}
\usepackage{comment}
\usepackage{makecell}

\usepackage[pagebackref=true,breaklinks=true,letterpaper=true,colorlinks,bookmarks=false]{hyperref}
\usepackage[capitalize]{cleveref}

\makeatletter
\DeclareRobustCommand\onedot{\futurelet\@let@token\@onedot}
\def\@onedot{\ifx\@let@token.\else.\null\fi\xspace}

\def\eg{\emph{e.g}\onedot} 
\def\ie{\emph{i.e}\onedot}

\makeatother
\newcommand{\mbf}{\boldsymbol}

\newcommand{\tablestyle}[2]{\setlength{\tabcolsep}{#1}\renewcommand{\arraystretch}{#2}\centering\small}

\crefname{section}{Sec.}{Secs.}
\Crefname{section}{Section}{Sections}
\Crefname{table}{Table}{Tables}
\crefname{table}{Tab.}{Tabs.}

\iccvfinalcopy 

\begin{document}

\title{CancerUniT: Towards a Single Unified Model for Effective Detection, Segmentation, and Diagnosis of Eight Major Cancers Using a Large Collection of CT Scans}

\author{
	Jieneng Chen$^{1,2}$\footnotemark[1]
	\;\; Yingda Xia$^{1}$\footnotemark[1]
	\;\; Jiawen Yao$^{1,3}$
        \;\; Ke Yan$^{1,3}$
        \;\; Jianpeng Zhang$^{1,3}$
        \;\; Le Lu$^1$
        \;\; Fakai Wang$^{1}$ \\
        \;\; Bo Zhou$^{1}$
        \;\; Mingyan Qiu$^{1,3}$ 
        \;\; Qihang Yu$^{2}$
        \;\; Mingze Yuan$^{1,3}$
        \;\; Wei Fang$^{1,3}$
        \;\; Yuxing Tang $^1$
        \;\; Minfeng Xu$^{1,3}$ \\
        \;\; Jian Zhou$^4$ 
        \;\; Yuqian Zhao$^5$ 
        \;\; Qifeng Wang$^5$ 
        \;\; Xianghua Ye$^6$ 
        \;\; Xiaoli Yin$^7$ 
        \;\; Yu Shi$^7$ 
        \;\; Xin Chen$^{8,9}$ \\
        \;\; Jingren Zhou$^{1,3}$
        \;\; Alan Yuille$^2$
	\;\; Zaiyi Liu$^{8,9}$\footnotemark[1]
        \;\; Ling Zhang$^1$  \\
	$^1$DAMO Academy, Alibaba Group \;\;  $^2$Johns Hopkins University \;\;  \\
 $^3$ Hupan Lab, 310023, Hangzhou, China \;\;
 $^4$Sun Yat-sen University Cancer Center \;\;  \\
 $^5$Sichuan Cancer Hospital \;\;
$^6$The First Affiliated Hospital of Zhejiang University \;\; \\
$^7$ Shengjing Hospital of China Medical University \;\; 
  $^8$Guangdong Provincial People’s Hospital \;\; \\
$^9$ Guangdong Key Laboratory of Artificial Intelligence in Medical Image Analysis and Application
}

\maketitle

\renewcommand{\thefootnote}{\fnsymbol{footnote}}
\footnotetext[1]{Correspondence to Jieneng Chen (jienengchen01@gmail.com), Yingda Xia (yingda.xia@alibaba-inc.com), and Zaiyi Liu (zyliu@163.com)}

\begin{abstract} \vspace{-3mm}
Human readers or radiologists routinely perform full-body multi-organ multi-disease detection and diagnosis in clinical practice, while most medical AI systems are built to focus on single organs with a narrow list of a few diseases. This might severely limit AI's clinical adoption. A certain number of AI models need to be assembled non-trivially to match the diagnostic process of a human reading a CT scan. In this paper, we construct a Unified Tumor Transformer (CancerUniT) model to jointly detect tumor existence \& location and diagnose tumor characteristics for eight major cancers in CT scans. CancerUniT is a query-based Mask Transformer model with the output of multi-tumor prediction. We decouple the object queries into organ queries, tumor detection queries and tumor diagnosis queries, and further establish hierarchical relationships among the three groups. This clinically-inspired architecture effectively assists inter- and intra-organ representation learning of tumors and facilitates the resolution of these complex, anatomically related multi-organ cancer image reading tasks. CancerUniT is trained end-to-end using a curated large-scale CT images of 10,042 patients including eight major types of cancers and occurring non-cancer tumors (all are pathology-confirmed with 3D tumor masks annotated by radiologists). On the test set of 631 patients, CancerUniT has demonstrated strong performance under a set of clinically relevant evaluation metrics, substantially outperforming both multi-disease methods and an assembly of eight single-organ expert models in tumor detection, segmentation, and diagnosis. This moves one step closer towards a universal high performance cancer screening tool. 

\end{abstract}
\section{Introduction}
\vspace{-0.1cm}
\begin{figure}[!tbp]
\centering
\includegraphics[width=0.9\linewidth]{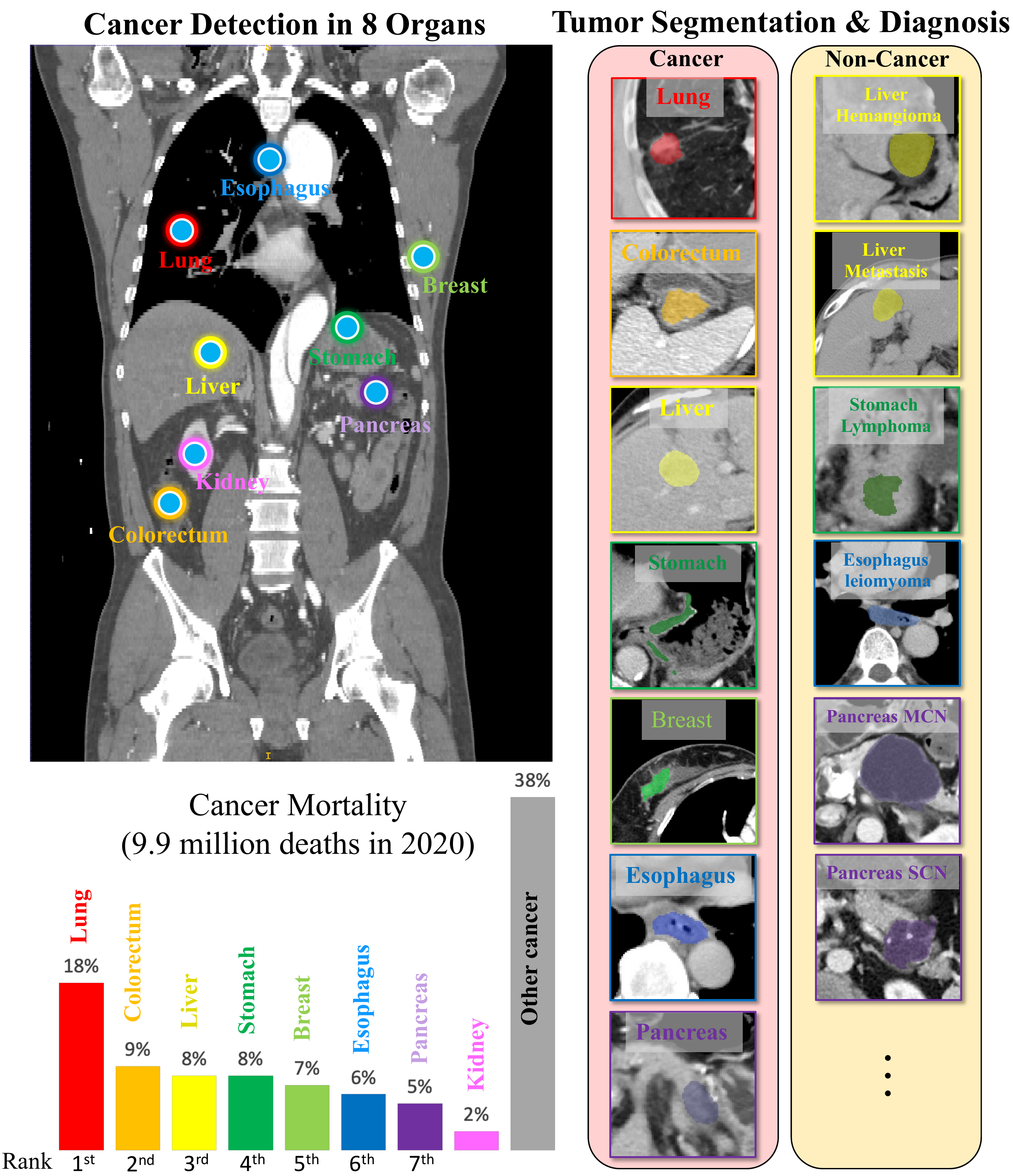}
\caption{We aim at cancer and non-cancer detection, segmentation, and diagnosis in eight major organs via CT scan. Seven of our eight targeted cancers rank the top seven in terms of mortality. }
\label{fig:aim}
\end{figure}

Cancer, a leading cause of death in the world, continues to thwart human life expectancy and cause huge societal burdens despite significant progress in medical research \cite{who2020global,sung2021global}. Medical imaging is a powerful tool for detection and diagnostic examination of cancer and is widely used in clinical practice everywhere. The daily work of radiologists in reading and interpreting cancer imaging findings includes three main clinical tasks: detection, quantification, and diagnosis \cite{bi2019artificial}. Since  Computed Tomography (CT) body scans are very common (nearly 80\% in all CT exams) \cite{sodickson2009recurrent} and each CT scan can have hundreds of image slices, the miss-detection and miss-diagnosis of cancer are the pain points in the radiology workflow. Human readers statistically tend to have high specificity but low sensitivity for tagging and reporting various anomalies or diseases.

Computer-aided detection (CADe) and diagnosis (CADx) can assist radiologists and oncologists to improve the tumor detection rate and diagnosis accuracy~\cite{doi2007computer,bi2019artificial}. With the development of deep learning and Convolutional Neural Networks (CNNs), CAD algorithms have met or exceeded expert-level performance in some specific applications \cite{mckinney2020international, liu2020deep, de2018clinically_nat2, ardila2019end_nat3}. However, most CAD expert systems focus on dealing with single organ diseases \cite{van2021artificial}, e.g., pancreas \cite{zhu2019multi,zhao20213d}, liver \cite{huo2020harvesting,cheng2022flexible}, lung \cite{ardila2019end_nat3,kim2022artificial}, or kidney tumors \cite{heller2021state_kits}, while radiologists in turn, must be responsible for all possible diseases and radiologically significant findings \cite{pickhardt2022value}. For example, for an abdominal CT examination that is initially targeted for the gallbladder, even when there are no clear prior indications (e.g., abdominal pain), all visible organs in the entire abdomen from CT imaging need to be carefully inspected by a radiologist. Therefore, the role of current CAD tools is still (very) limited in clinical practice, far from functioning as universally as a human reader. A versatile CAD tool that can perform (many) more critical medical tasks would be more clinically desirable and thus is in high demand \cite{summers2003road}. 

Despite notable progress in multi-organ segmentation, it should be noted that detecting and diagnosing multiple cancers is considerably more difficult than segmenting organs alone due to several factors: (1) tumors have a variety of types, appearances and size, making it hard to be detected. (2) tumor detection requires differentiation of tumors from normal tissue within an organ, which is more challenging than the differentiation of organs from the background in organ segmentation. (3) the diagnosis of cancer involves the fine-grained categorization of tumors, which necessitates a high level of expertise and specialized training. Aiming at solving the universal lesion detection problem in CT scans, DeepLesion is a recent pioneering publicly available dataset~\cite{yan2018deep,yan2018deeplesion,yan2019holistic}, and despite much follow-up work, most cancer detection, quantification, and diagnosis solutions derived from DeepLesion dataset ~\cite{yan2018deep} are still insufficient in the following aspects. First, the data size and patient number of a single disease can be small, and some major cancer types (e.g., esophagus, stomach, and colorectum) are scarce, resulting in relatively high false positive and sub-optimal detection rates. Second, voxel-level tumor annotation, perhaps as 3D masks (requiring a high level of clinical expertise and are very tedious to label) are not available, making the necessary precise 3D quantification difficult (if not impossible). Third, the pathological gold standard of confirming tumor types is unavailable and, therefore, impossible to distinguish between malignant and benign lesions. Recent clinical validations of two multi-disease detection AI systems \cite{rueckel2021reduction,xu2021external} found that ruling out irrelevant CAD findings (i.e., false positives and lesions without adequate malignancy assessment) was very time-consuming and confusing for radiologists. These observations clearly indicates the essential limitations of applying DeepLesion dataset ~\cite{yan2018deep} to positive clinical impacts.

In this paper, we curate a large (abdominal and chest) CT image dataset including eight major cancers (from the top seven cancers with the highest mortality in the world \cite{sung2021global}: lung, colorectum, liver, stomach, breast, esophagus, and pancreas, plus a public kidney dataset \cite{heller2021state_kits}) of total 10,673 patients (Of these, breast cancer has the fewest, with 478 patients; lung cancer has the most, with 2,402 patients. In addition, there are 1,055 normal controls). All tumor types (and subtypes) of the seven organs are confirmed by either surgical or biopsy pathology and recorded as gold standard labels, where full spectrum of all tumor subtypes are offered for four organs. All confirmed tumors in CT scans are manually segmented or delineated in 3D by board-certified radiologists who are specialized in the particular organ or disease types. To our knowledge, previous datasets with similar tumor characteristics only cover a single disease at the scale of hundreds of patients, such as the pancreatic tumor \cite{zhao20213d} and kidney tumor datasets \cite{heller2021state_kits}. 
The curation of our new 8-cancer dataset is a major step towards building a universal multi-cancer imaging reading AI model -- with the hope to reach a performance level comparable with radiologists specializing in different cancer types -- for assisting radiologists and general clinicians in precision detection, quantification, and diagnosis. Fig.~\ref{fig:aim} shows our goal for cancer and non-cancer detection, segmentation, and diagnosis in eight major organs via CT scans.

On the other hand, we propose a new clinically interpretable computing architecture, named Unified Tumor Transformer (CancerUniT). In general, CancerUniT is a single unified model that simultaneously solves the tasks of multi-tumor detection, segmentation and diagnosis in a semantic segmentation manner. Our motivations are: (1) the organs, cancers and non-cancer tumors are interrelated in both appearance similarity and human anatomical constraints, e.g., HCC (a major malignant liver cancer) and cyst (benign lesion) both occur inside the liver with textual and other visual differences, while HCC and PDAC (a type of pancreatic cancer) should appear in two different organs but their clinical characteristics are both malignant carcinoma; (2) a unified learning of multi-organs-tumors could reduce the performance uncertainty and architectural complexity in assembling multiple single models, e.g., different predictions of the same intended object or finding by multiple models. To collaboratively model such differences and connections or dependencies, we propose a novel representation learning method that represents each organ and tumor as an object query of the Transformer in a semantic hierarchy. The object queries are divided into organ queries, tumor detection queries and diagnosis queries, and we establish a query hierarchy based on the clinical meaning of the queries. This design will explicitly encourage the queries to learn the inter-organ and intra-organ relationships to solve the clinically sophisticated multi-cancer tumor recognition tasks.

CancerUniT is trained and tested on our curated dataset.  
CancerUniT outperforms the DeepLesion model, the ensemble of single-organ expert models and unified baseline models (trained on our data). Compared to the DeepLesion model, CancerUniT has a 29.3\% higher sensitivity and a large margin of 77.5\% higher specificity in tumor detection. Compared to an ensemble of individually trained single-organ nnUNet models, CancerUniT has an average improvement of 6.7\% in tumor detection sensitivity, 2.8\% in diagnostic accuracy, and 3.9\% in Dice segmentation score across all the organs; On normal patients, CancerUniT has an improvement of 22.5\% in specificity (ours 81.7\% vs. nnUNet 59.2\%); CancerUniT is 4.5 times faster in testing speed. In comparison to a unified nnUNet model dealing with all eight organs, CancerUniT leads by 5.3\% in lesion detection sensitivity, 6.7\% in diagnostic accuracy, 2.8\% in specificity, and 2.7\% in Dice segmentation score. The improvements indicate that the different type of tumors have mutual correlations and the design of CancerUniT successfully captures this clinical relationship for enhanced tumor representation learning.
The high performance of CancerUniT also sheds light on its clinical potential for real-world multi-cancer detection, segmentation, and diagnosis. 

\section{Related Work}


{\bf CADe and CADx.} CADe normally refers to the computer-aided localization process of lesions in 2D/3D medical images and CADx subsequently diagnoses lesions or findings as either malignant or benign \cite{bi2019artificial} and assigning more potential tumor characteristics. 
Along with advances in deep learning, quantitative CADe performances matching or beyond medical domain experts are reported in several specialized single-organ clinical applications: breast cancer screening~\cite{mckinney2020international}, lung cancer detection \cite{ardila2019end_nat3}, retinal disease referral \cite{de2018clinically_nat2}, skin disease diagnosis \cite{liu2020deep} and so on.

{\bf Tumor detection, segmentation, and diagnosis in CT via CNNs.} CNNs have been widely applied to detect, segment, and diagnose cancers/tumors in CT scans. Lung nodule detection in low-dose CT~\cite{national2011national} is the recommended lung cancer screening protocol where some promising results are discussed~\cite{ardila2019end_nat3,huang2017lung_lung1,xie2018knowledge_lung2,zheng2019automatic_lung3}. Image segmentation networks \cite{long2015fully, chen2017deeplab, ronneberger2015u, isensee2021nnu, zhou2019unet++} are well-adopted under the per-pixel classification setting and a segmentation model is designed to predict the probability distribution over all possible categories or labels per pixel (as a structured dense prediction problem). Segmenting abdominal organs and detecting tumor by segmentation principles \cite{wasserthal2022totalsegmentator,antonelli2022medical,bilic2019liver,heller2021state_kits,zhang2021modality,yao2022deepcrc,hosny2022clinical},  
serve a key role towards fully-automated tumor detection~\cite{zhu2019multi,xia2021effective,yao2022effective, xia2022felix}, differential diagnosis and reporting~\cite{zhao20213d}. Despite their promising performance, these approaches are often specialized to focus only on a single organ. 
Multi-organ segmentation~\cite{luo2022word, lee2020semi, wang2019abdominal, zhang2021dodnet, dmitriev2019learning, zhou2019prior}
 are emerging, but the degree of difficulty involved in multi-cancer detection and diagnosis is considerably greater than that of organ-level. DeepLesion~\cite{yan2018deep} attempts to tackle the universal lesion detection task in CT scans, but their derived lesion detection methods ~\cite{yan2018deeplesion,yan2019holistic,yan2021learning} and several follow-up work \cite{yang2021asymmetric,Lyu2021Segmentation,yan2019MULAN, tang2019uldor, tang2021weakly} so far have reported mostly moderate multi-class lesion detection performance. Distinguishing between malignant and benign lesions in multi-class tumor setting is still far from a clinical reality. 


{\bf Transformers} ~\cite{vaswani2017attention} have advanced the state-of-the-art performance in various computer vision tasks~\cite{dosovitskiy2020image, carion2020end, liu2021swin, chen2022transmix, fan2021multiscale, zhu2020deformable, touvron2021training}, by capturing global interactions between image patches and having no built-in inductive prior. The success of Transformer has also been witnessed in medical image detection and segmentation~\cite{chen2021transunet, xie2021cotr, hatamizadeh2022swin}. With the recent progress in transformers \cite{carion2020end, wang2021max}, a new variant called mask Transformers has been proposed, where segmentation predictions are represented by a set of query embeddings with their own semantic class labels, generated through the conversion of query embedding to mask embedding vectors followed by multiplying with the image features. The essential component of mask transformers is the decoder which takes object queries as input and gradually transfers them into mask embedding vectors.  Recent works~\cite{strudel2021segmenter, wang2021max, cheng2021per, cheng2022masked} inspire us to represent tumor in the medical domain as the class query~\cite{strudel2021segmenter} within the Transformer formulation. In this paper, we propose a novel semantic hierarchical representation to exploit the relationship in detection, diagnosis and differentiation among eight main tumors and their sub-types from a large dataset of CT scans collected from both healthy subjects and patients with cancers. 


\section{Method}
In this section, we first define the problem of tumor detection, segmentation, and diagnosis from an image semantic segmentation perspective in Sec.~\ref{sec:pro-def}. We then describe the overview of query-based mask Transformer and how we integrate it as our segmentation decoder in Sec.~\ref{sec:mask}. After that, we introduce the proposed Unified tumor Transformer (CancerUniT) in Sec.~\ref{sec:unit}, which represents tumors by a semantic query hierarchy, solving the tumor detection, segmentation, and diagnosis in a unified manner. 

\subsection{Problem Definition}
\label{sec:pro-def}
 We focus on three tasks in images, i.e., tumor detection, segmentation, and diagnosis. Tumor detection aims to locate the presence of target types of tumors. Tumor segmentation aims to provide per-pixel annotation of the tumor region. Tumor diagnosis aims to classify the specific tumor subtype of a detected tumor. We denote $\mathbf{s}_{o}$ as a set of organs, and $\mathbf{s}_{t}$ as a set of tumors. Specifically in our dataset,

\begin{table}[htb]
    \centering
    \scriptsize
    \begin{tabular}{l|l}
    \hline
      $\mathbf{s}_{o}$   &  \{breast, lung, kidney, pancreas, esophagus, liver, stomach, colorectum\} \\ \hline
      $\mathbf{s}_{t}$   & \thead{\scriptsize \{breast cancer, lung cancer, colorectal cancer, \\ \scriptsize pancreas PDAC, pancreas nonPDAC, liver HCC, liver ICC, \\  \scriptsize liver metastatis, liver hemangioma,  stomach GC, \\ \scriptsize stomach nonGC,  esophagus EC,  esophagus nonEC, \\  \scriptsize kidney tumor/cyst\}} \\

     \hline
    \end{tabular}
    \label{tab:my_label}
\end{table}

We propose to solve these three tasks with a semantic segmentation framework, in which we assign each voxel in the CT scan with a semantic label $k \in \mathbf{s}_{o} \cup \mathbf{s}_{t}$ and the total number of classes is $K=|\mathbf{s}_{o}| + |\mathbf{s}_{t}| $ . The tumor detection, segmentation, and diagnosis are then evaluated based on the semantic segmentation results.

\subsection{Basis: Query-based Mask Transformers}
\label{sec:mask}
Although Transformers have been used for medical image segmentation as feature extractors, the query-based mask Transformer~\cite{strudel2021segmenter, wang2021max, cheng2021per} decoder is rarely explored in medical images. Query-based mask Transformer aims to decode the pixel-level features (usually from a CNN backbone) with object queries. Our method is based on this design and here we provide an overview of its basic components.
\textbf{Query initialization.} 
A set of $K$ learnable class queries (\ie, embeddings) $\mathbf{q} = [\text{q}_1, ..., \text{q}_K] \in \mathbb{R}^{K \times d}$ is defined where $K$ is the number of classes and $d$ is the query dimension. Each class query is initialized randomly and assigned to a single semantic class.

\textbf{Query interaction via Transformer.}
The queries are updated through multi-head cross-attention, multi-head self-attention, and feedforward network~\cite{vaswani2017attention}. The multi-head cross-attention between queries and image features is computed to update queries conditioned on image features. The multi-head self-attention allows queries to interact with each other.

\textbf{Decode queries to segmentation.}
The class query $\mathbf{q}$ is processed jointly with 3D image features $\mathbf{F}  \in \mathbb{R}^{d \times D \times H \times W}$ by the decoder. 
$K$ masks can be generated by computing the scalar product between L2-normalized image features  $\mathbf{F} \in \mathbb{R}^{d \times D \times H \times W}$ and class queries $\mathbf{q} \in \mathbb{R}^{K\times d}$.  The set of class masks is computed as:
\begin{equation}
  \mathbf{M} = \mathbf{q}\times \mathbf{F}
  \label{equ:maskdecode}
\end{equation}
where $\mathbf{M} \in \mathbb{R}^{K \times D \times H \times W}$ 
is $K$ mask predictions and will be followed by a softmax to obtain the final pixel-wise class probability map in the task of semantic segmentation. 

\subsection{CancerUniT: Unified Tumor Transformer}
\label{sec:unit}
\begin{figure}[!tbp]
\centering
\includegraphics[width=0.9\linewidth]{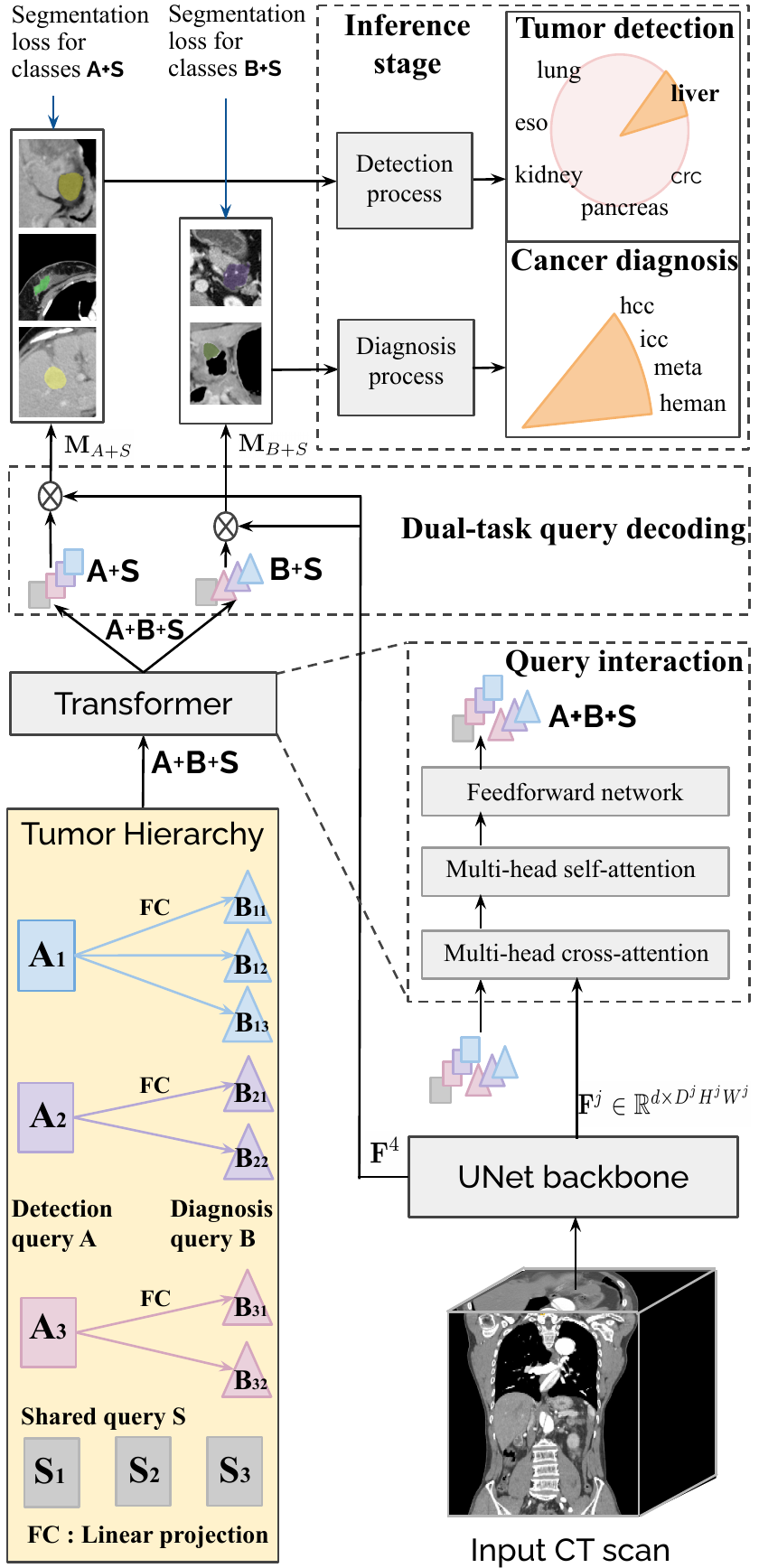}
\caption{Overview of Unified Tumor Transformer (CancerUniT). We first represent tumor as queries $\mathbf{A}$ and $\mathbf{B}$ (\ie feature embedding), and then build the query hierarchy from  $\mathbf{A}$ to  $\mathbf{B}$ via linear projection $\mathbf{FC}$ according to the tumor sub-type relationship. The tumor queries interact and are updated in a Transformer decoder with input of UNet features $\mathbf{F}$. Dual-task query decoding is performed to generate semantic segmentation map for two tasks. The detection task focuses on major tumor classes while diagnosis task is supervised by fine-grained cancer sub-types. In inference stage, the dual-task tumor segmentation maps are post-processed separately to produce multi-classes tumor instances for tumor detection and cancer diagnosis.}

\vspace{-0.3cm}
\label{fig:arch}
\end{figure}

We introduce the novel unified tumor transformer (See Fig.~\ref{fig:arch}), which includes semantic query hierarchy for tumor representation, UNet backbone for feature extraction, Transformer for query interaction, and dual-task query decoding for tumor detection task and cancer diagnosis task. 

\vspace{-0.2cm}

\subsubsection{Query Hierarchy}

\label{hierarchy}
We propose a novel tumor representation via a semantic hierarchy of queries, including shared, detection, and diagnosis queries. 
By this design, tumors are represented as queries and a ``detection-to-diagnosis hierarchy" is established based on the semantic relationship of tumors. 

We hereby divide the segmentation targets $\mathbf{s}_o \cup \mathbf{s}_t$ into three non-overlapping groups, i.e., $\mathbf{m}$, $\mathbf{n}$, and  $\mathbf{s}$. $\mathbf{m}$ consists of $m$ general tumor categories that requires further diagnosis.  
The $i^{th}$ element $\mathbf{m}_i$ can be further categorized into the sub-classes of $\mathbf{n}_i$ with a number of $n_i$ sub-classes. $\mathbf{s}$ consists the rest of the targets including eight organs and the four cancers that do not require diagnosis in our data.

\textbf{Shared query.}   We define a set of shared query $\mathbf{s} \in \mathbb{R}^{s\times d}$ to represent the shared classes in both detection task and diagnosis task. The shared classes include the 8 organ classes, and 4 tumor classes without other sub-types.

\begin{table}[htb]
    \centering
    \scriptsize
    \begin{tabular}{l|l}
    \hline
      $\mathbf{s}$   &  $\mathbf{s}_o$ $\cup$ \{lung cancer, breast cancer, colorectal cancer, kidney tumor/cyst\} \\ \hline
      $\mathbf{m}$   & \{pancreas tumor, liver tumor, stomach tumor, esophagus tumor\} \\ \hline
      $\mathbf{n}$ & \thead{\scriptsize \{\{PDAC, nonPDAC\}, \{HCC, ICC, metastasis, hemangioma\}, \\ 
      \scriptsize \{GC, nonGC\}, \{EC, nonEC)\}\}}  \\
\hline
    \end{tabular}
    \label{tab:my_label}
    \vspace{-0.3cm}
\end{table}

\textbf{Detection query.} We denote $\mathbf{A} \in \mathbb{R}^{m\times d}$ as detection queries with $m$ specifying the number of queries. Each detection query corresponds to the general tumor class of an organ, which requires further diagnosis. 

\textbf{Diagnosis query.}  
The cancer diagnosis relies on fine-grained tumor categorization. Similarly, we denote a feature embedding $\mathbf{B} \in \mathbb{R}^{n_i\times d}$ as diagnosis queries with $n_i$ specifying the number of diagnosis classes for tumor $\mathbf{m_i}$. A group of $n_i$ diagnosis queries refers to $|\mathbf{n_i}|$ tumor sub-types $\mathbf{n_i}$ occurred in organ $i$, and totally we have $n = \sum_{i=1}^{|\mathbf{n}|}n_i$ diagnosis queries in this work. 

\textbf{Detection-to-Diagnosis hierarchy via linear projection.} Inspired by the clinical practice of detection-then-diagnosis and given the fact that diagnosis queries are subtypes of detection queries, we aim to build a graph treating the detection queries as parent nodes and diagnosis queries as children nodes. In this way, the model is able to learn the hierarchical representation of tumors explicitly.

To build the semantic hierarchical relationship, we project a detection query $\mathbf{A}_i\in\mathbb{R}^{1 \times d}$ into diagnosis queries $\mathbf{B}_i \in \mathbb{R}^{1 \times n_i d}$ via a linear projection layer with matrix $\mathbf{W}_i \in \mathbb{R}^{n_i d \times d}$. 
The detection-to-diagnosis procedure is formulated as:
\vspace{-0.1cm}
\begin{equation}
  \mathbf{B}_i = \mathbf{A}_i \times \mathbf{W}_i^T
\end{equation} where $\mathbf{B}_i = [\mathbf{B}_{i,1}, \mathbf{B}_{i,2}, ... \mathbf{B}_{i,n_i}]$ and the subgroup capacity $n_i = |\mathbf{n_i}|$ represents the number of subtypes for tumor $\mathbf{m_i}$. The detection queries $\mathbf{A}$ are learnable parameters that are random initialized, while diagnosis queries $\mathbf{B}$ are feature embedding conditioned on detection queries.

\subsubsection{Meta Architecture}
The proposed architecture includes a UNet backbone for feature extraction, a Transformer for query interaction, and a dual-task query decoding stage to generate segmentation masks. Detailed model instantiation is in Appendix.

\textbf{nnUNet backbone for feature extraction.} 
We adopt nnUNet~\cite{isensee2021nnu} as the backbone to extract multi-scale features $\mathbf{F}=[\mathbf{F}^1, \mathbf{F}^2, \mathbf{F}^3, \mathbf{F}^4]$ where  
$\mathbf{F}^j\in \mathbb{R}^{d \times D^j H^j W^j}$ 
is the $j$-th layer feature map after projecting to number of channel $d$ and flattening the spatial dimension $D^j$, $H^j$, and $W^j$. 

\textbf{Transformer for query interaction.}
We use the standard Transformer decoder~\cite{vaswani2017attention} with input of UNet features $\mathbf{F}^j$ and queries $[\mathbf{A}^j, \mathbf{B}^j, \mathbf{S}^j]$ at the $j$-th layer. The Transformer is stacked by three Transformer layers, each of which contains a multi-head cross-attention, a multi-head self-attention, and a feed-forward network. The concatenated query $[\mathbf{A}, \mathbf{B}, \mathbf{S}]$ is updated via the cross attention (denoted as $CA$) between the queries and the image feature $\mathbf{F}^j$, as well as the query self-attention (denoted as $SA$).
The query interaction is written as:
\vspace{-0.2cm}

\begin{align}
\begin{split}
\mathbf{A}^j, \mathbf{B}^j, \mathbf{S}^j &= SA(CA([\mathbf{A}^{j-1}, \mathbf{B}^{j-1}, \mathbf{S}^{j-1}], \mathbf{F}^j))
\end{split}
\end{align}

\textbf{Dual-task query decoding.}
As there exists inclusiveness for the classes in $\mathbf{A}$ and $\mathbf{B}$, it is unlikely to decode them jointly if we would like to enforce multi-class exclusivity constraint (\eg softmax).  To better capture the class exclusivity, we propose the dual-task query decoding procedure that decodes queries $[\mathbf{A}, \mathbf{S}]$ and queries $[\mathbf{B}, \mathbf{S}]$ separately to perform dual-task semantic segmentation. The query decoding follows Eq.~\ref{equ:maskdecode} with a softmax activation, written as: 
\vspace{-0.4cm}
\begin{align}
\begin{split}
\mathbf{M}_{A+S} = softmax([\mathbf{A}, \mathbf{S}] \times \mathbf{F}^4) \\
\mathbf{M}_{B+S} = softmax([\mathbf{B}, \mathbf{S}] \times \mathbf{F}^4) 
\end{split}
\end{align} 
where $\mathbf{M}_{A+S}$ and $\mathbf{M}_{B+S}$ are the decoded voxel-wise semantic map for the detection task and the diagnosis task, respectively.

\textbf{End-to-end training.}
Our method performs both major tumor segmentation and tumor subtype segmentation directly from CT scans, while vanilla methods only output subtype segmentation maps that are further merged to major tumor segmentation maps. In our work, the loss function 
is the combination of cross-entropy loss and Dice loss~\cite{milletari2016v}, which are applied to both detection output $\mathbf{M}_{A+S}$ and diagnosis output $\mathbf{M}_{B+S}$ to enforce the similarity with their corresponding targets.

\textbf{Inference.}
End-to-end inference of dual-task segmentation is enabled simultaneously. For the detection process, the tumor segmentation map from $\mathbf{M}_{A+S}$ is extracted to generate tumor instances (\ie connected components) with tumor class label $i$. If the predicted tumor instance
has overlaps with the ground-truth tumor, the patient is detected with tumor class $i$. For the diagnosis process, we do similar tumor instance extraction from $\mathbf{M}_{B+S}$, but each tumor instance is identified as one specific tumor subtype. 
The patient-level cancer diagnosis category is decided by the tumor 
subtype with the largest connected component.

\begin{table*}[]
\scriptsize
\setlength{\tabcolsep}{1.5mm}
  \centering
    \caption{Dataset description. A-F denote six different hospitals.}
\begin{tabular}{c|cccc|cccccccccc|cc|c}
\toprule
\multirow{3}{*}{ }  & \multicolumn{4}{c|}{Cancers} & \multicolumn{10}{c|}{Full spectrum tumors} & \multicolumn{2}{c|}{Normal controls } & \multirow{2}{*}{Total }\\
 & Breast & CRC  & Kidney  & Lung  & \multicolumn{2}{c|}{Pancreas}  & \multicolumn{2}{c|}{Esophagus} & \multicolumn{2}{c|}{Stomach} & \multicolumn{4}{c|}{Liver} & Abdomen  & CTA  &  \\ \hline
 Hospitals & A & A  & public  & A,B  & \multicolumn{2}{c|}{C}  & \multicolumn{2}{c|}{B,D,E} & \multicolumn{2}{c|}{A} & \multicolumn{4}{c|}{C} & A  & F  &  \\ \hline
Subtypes & BC & CRC  & KT  & LC  & PDAC & \multicolumn{1}{c|}{nonPDAC}  & EC & \multicolumn{1}{c|}{nonEC} & GC & \multicolumn{1}{c|}{nonGC} & HCC & ICC & Meta & Heman & -  & - & \\ \hline
\begin{tabular}[c]{@{}c@{}}Train\end{tabular} & 428 & 746  & 249  & 2352  & 1315 & \multicolumn{1}{c|}{727}  & 1185 & \multicolumn{1}{c|}{105} & 1117 & \multicolumn{1}{c|}{273} & 284 & 31 & 99 & 147 & 884  & 100  & 10042    \\
\begin{tabular}[c]{@{}c@{}}Test\end{tabular} & 50 & 50  & 50  & 50  & 50 & \multicolumn{1}{c|}{50}  & 50 & \multicolumn{1}{c|}{50} & 50 & \multicolumn{1}{c|}{50} & 15 & 15 & 15 & 15 & 50  & 21  & 631    \\ \hline
\begin{tabular}[c]{@{}c@{}}Total\end{tabular} & 478 & 796  & 299  & 2402  & 1365 & \multicolumn{1}{c|}{777}  & 1235 & \multicolumn{1}{c|}{155} & 1167 & \multicolumn{1}{c|}{323} & 299 & 46 & 114 & 162 & 934  & 121  & 10673    \\
\toprule
\end{tabular}
\label{tab:data}
\vspace{-0.3cm}
\end{table*}

\section{Experiments}
\subsection{Experiment Setup}

\textbf{Dataset description.}
Our 8-cancer CT dataset, which includes seven in-house tumor datasets (collected from five hospitals), one publicly available kidney tumor dataset \cite{heller2021state_kits}, and a normal control dataset, is composed of 10,673 contrast-enhanced CT volumes (all in venous phase, except for lung and CT angiography being arterial phase), each from one unique patient. 
These CT volumes are acquired before treatment.
All cancers (and tumor subtypes) in the seven in-house datasets are confirmed by pathology, with four datasets having a full spectrum of tumor subtypes, i.e., liver (4 subtypes), stomach (6), esophagus (4), and pancreas (9). The normal controls consist of 934 abdominal CT and 121 CT angiography (CTA) scans. 
Some of the datasets for single organs have been involved in our previous publications for other precision oncology research purposes \cite{zhao20213d,yao2022deep,yao2022effective,yao2022deepcrc,huang2016development,wang2022mining,dong2020deep,yang2020deep}.

 Tumors in each organ dataset are manually segmented slice-by-slice on CT images by radiologists who provide the data and specialize in the specific disease using either ITK-SNAP \cite{py06nimg} or our in-house developed CT  annotation tool -- CTLabler \cite{wang2023multi}. During annotation, the radiologists also refer to the other CT phases (e.g., arterial, delay), contrast-enhanced MRI, and radiology/surgery/pathology reports if necessary.
All organs are segmented/delineated automatically: the breast is by a nnUNet model trained on 213 additional breast cancer CT volumes with CTV (clinical target volume) masks; the other seven organs are by another nnUNet model trained on the Totalsegmentator dataset \cite{wasserthal2022totalsegmentator}. 
We randomly select 50 CT volumes from each tumor subtype (except for the liver tumor subtypes being 15 each, considering the relatively small liver data size), 50 abdominal, and 21 CTA volumes to form the test set. The remaining 10,042 CT volumes are used as the training set (Table \ref{tab:data}).

\textbf{Implementation details.}
All images were resampled to a spacing of $3\times0.8\times0.8$mm $(Z\times X\times Y)$. 
In the training stage, we randomly cropped sub-volumes of size of $48\times 192\times 192$ voxels
from CT scans as the input. 
We employed the online data augmentation of nnUNet ~\cite{isensee2021nnu}, including random rotation, scaling, flipping, Gaussian blurring, adding white Gaussian noise, adjusting brightness and contrast, simulation of low resolution, and Gamma transformation, to diversify the training set. The balanced sampling strategy was adopted to encourage model to sample different datasets and also different organ regions evenly. The batch size was set to 8, with 1 batch size per GPU on an 8-GPU machine. We adopted the AdamW optimizer and an initial learning rate of 3e-4. The baseline models were trained from scratch with 700 epochs, and the number of iterations per epoch equaled to training dataset size divided by the batch size. It took 40 GPU days to train a nnUNet from scratch on our dataset with Nvidia V100 GPUs. Due to huge cost, CancerUniT was trained based on the pre-trained nnUNet with a learning rate multiplier 0.1, and we trained 50 epochs. 
For fair comparison, we also kept tuning nnUNet for another 50 epochs besides 700 epochs whereas no performance improvement was observed.  

In the inference stage, we employed the sliding window strategy, where the window size equals to the training patch size. In addition, Gaussian importance weighting and test time augmentation by flipping along all axes were also utilized to improve the robustness of segmentation.


\begin{table*}[ht!]
\small
  \centering
    \caption{Patient-level tumor detection results. Sensitivity (\%) and specificity (\%) are reported. ``DeepLesion model'' is trained on DeepLesion dataset~\cite{yan2018deeplesion}) using the detection-based algorithm LENS~\cite{yan2021learning}; ``LENS (trained on our data)'' is trained on our new 8-cancer dataset.}
\begin{tabular}{c|ccccccccc|ccc}
\toprule
\multirow{2}{*}{Model}                                      & \multicolumn{9}{c|}{Sensitivity (\%)}                                & \multicolumn{3}{c}{Specificity (\%) } \\
                                                            & Br. & Crc.  & Kid.  & Lung  & Pan.  & Eso.  & St.  & Liv.  & Average & Abd.   & CTA   & Average   \\ \hline
\begin{tabular}[c]{@{}c@{}}8-nnUNet ensemble\end{tabular} & \textbf{96.0} & 74.0 & 94.0 & 74.0 & 93.0 & 83.0 & 92.0 & 86.7 & 86.6    & 80.0      & 9.5     & 59.2      \\
DeepLesion model~\cite{yan2018deep}                                                   & 78.0 & 38.0 & 86.0 & 76.0 & 82.0 & 34.0 & 30.0 & 88.3 & 64.0    & 6.0       & 0.0     & 4.2       \\
LENS (train on our data)~\cite{yan2021learning}                                                        & 82.0 & 62.0 & 76.0 & 50.0 & 89.0 & 72.0 & 72.0 & 75.0 & 72.3    & 70.0      & 52.4    & 64.8      \\
nnUNet~\cite{isensee2021nnu}                                                       & 90.0 & 82.0 & 92.0 & 94.0 & 94.0 & 76.0 & 91.0 & 85.0 & 88.0    & 92.0      & 47.6    & 78.9      \\
TransUNet~\cite{chen2021transunet,isensee2021nnu}                                               & 94.0 & 86.0 & 94.0 & 94.0 & 94.0 & 81.0 & \textbf{94.0} & 88.3 & 90.7    & 94.0      & 47.6    & 80.3      \\ \hline
Ours                                                        & 94.0 & \textbf{92.0} & \textbf{94.0} & \textbf{94.0} & \textbf{95.0} & \textbf{89.0} & 93.0 & \textbf{95.0} & \textbf{93.3}    & \textbf{96.0}      & \textbf{47.6}    & \textbf{81.7}      \\ 
\toprule
\end{tabular}
\label{tab1}
\vspace{-0.3cm}
\end{table*}

\textbf{Evaluation metrics.}
We consider the evaluation metrics from three aspects, including patient-level, lesion-level, and dense-level metrics. For patient-level evaluation, 
sensitivity and specificity are computed. Lesion-level precision and recall (do not consider the tumor type) are computed based on connected component analysis of tumor predictions. Tumor segmentation accuracy is assessed by the Dice coefficient.

\textbf{Baselines.} We compare our method to five baselines. (i) 8-nnUNet ensemble: An ensemble of 8 separately trained nnUNet~\cite{isensee2021nnu} models. To solve overlapping tumor predictions, we extract the tumor connected components from 8 model predictions and merge them with the priority of tumor size. (ii) nnUNet: A unified nnUNet trained on our dataset as a multi-organ multi-tumor segmentation task. (iii) TransUNet: A leading Transformer model TransUNet~\cite{chen2021transunet} on medical image segmentation implemented in the nnUNet framework~\cite{isensee2021nnu} and with the same settings as (ii).  (iv) DeepLesion model: A universal lesion detection model~\cite{yan2021learning} trained on the DeepLesion dataset~\cite{yan2018deep}. (v) LENS (train on our data): A leading medical lesion detection algorithm LENS~\cite{yan2021learning} trained on our dataset.

For a fair comparison, all segmentation-based models adopt fair data augmentations following nnUNet ~\cite{isensee2021nnu} and the same training techniques, while LENS and DeepLesion as detection-based methods adopt augmentations and training techniques following LENS ~\cite{yan2021learning}. 
All the models are trained to be converged.

\subsection{Main Results}

\begin{table}[]
\small
  \centering
    \caption{Class-agnostic lesion instance-level detection results. We treat eight types of tumors as one class. The numbers of FN, TP, FP lesions, precision and recall are reported.  The total number of ground-truth lesions in the test set is 767. Note one patient might have several lesions in the 560 patients with tumors, and a ground-truth lesion might be matched with multiple TP components.}
\label{tab:lesion}
\tablestyle{2.4pt}{1.1}

\begin{tabular}{c|ccc|cc}
\toprule
Model  & FN  & TP  & FP   & Precision & Recall  \\ \hline
8-nnUNet ensemble     & 209 & 568 & 1060 & 34.9\%   & 72.8\% \\
DeepLesion model~\cite{yan2018deep}           & 376 & 649 & 5345 & 10.8\%   & 51.0\% \\
LENS (train on our data)~\cite{yan2021learning}              & 267 & 602 & 875  & 40.8\%   & 65.2\% \\
nnUNet~\cite{isensee2021nnu}                & 223 & 557 & 585  & 48.8\%   & 70.9\% \\ 
TransUNet~\cite{chen2021transunet,isensee2021nnu}                & 169 & 648 & 726  & 47.2\%   & \textbf{77.9\%} \\ \hline
Ours                  & 192 & 592 & 508  & \textbf{53.8\%}   & 75.0\% \\ 
\toprule

\end{tabular}
\vspace{-0.5cm}
\end{table}

\begin{table*}[]
\small
 \centering
     \caption{Voxel-level tumor semantic segmentation results (in Dice coefficient \%). 
     }
\begin{tabular}{c|cccccccc|c}
\toprule
Model             & \multicolumn{1}{c}{Breast} & \multicolumn{1}{c}{Colorectum} & \multicolumn{1}{c}{Kidney} & \multicolumn{1}{c}{Lung} & \multicolumn{1}{c}{Pancreas} & \multicolumn{1}{c}{Esophagus} & \multicolumn{1}{c}{Stomach} & \multicolumn{1}{c|}{Liver} & average \\ \hline
8-nnUNet ensemble & 0.623                   & 0.474                   & 0.728                   & 0.415                   & 0.690                   & \textbf{0.661}                   & 0.420                   & 0.703                    & 0.589   \\
nnUNet~\cite{isensee2021nnu}             & 0.661                   & 0.515                   & 0.736                   & \textbf{0.548}                   & 0.695                   & 0.597                   & 0.418                   & 0.676                    & 0.601   \\
TransUNet~\cite{chen2021transunet,isensee2021nnu}     & 0.700                   & 0.530                   & 0.738                   & 0.540                   & 0.700                   & 0.621                   & \textbf{0.444}                   & 0.691                    & 0.620   \\ \hline
Ours              & \textbf{0.702}                   & \textbf{0.533}                   & \textbf{0.739}                   & 0.515                   & \textbf{0.702}                   & 0.652                   & 0.435                   & \textbf{0.743}                    & \textbf{0.628}   \\ 
\toprule
\end{tabular}

\label{tab:seg_organ}
\end{table*}

\begin{table*}[]
\small
  \centering
    \caption{Patient-level cancer diagnosis. The sensitivity (\%) for each tumor subtype is reported. We categorize tumor subtypes as two classes of cancer and non-cancer tumors for pancreas, esophagus, and stomach datasets; and consider four major subtypes for liver dataset.}
\tablestyle{3pt}{1.1}
\begin{tabular}{c|ccc|ccc|ccc|ccccc|c}
\toprule
\multirow{2}{*}{Model} & \multicolumn{3}{c|}{Pancreas} & \multicolumn{3}{c|}{Esophagus}  & \multicolumn{3}{c|}{Stomach} & \multicolumn{5}{c|}{Liver}        & \multirow{2}{*}{Average} \\
                       & PDAC    & nonPDAC    & avg    & EC & nonEC & avg  & GC      & nonGC    & avg     & HCC  & ICC  & Meta & Heman & avg  &                          \\ \hline
8-nnUNet ensemble      & 88.0    & 74.0       & 81.0   & 92.0   & 32.0      & 62.0 & 94.0    & 28.0     & 61.0    & \textbf{80.0} & 60.0 & \textbf{46.7} & 86.7  & \textbf{68.3} & 68.1                     \\
nnUNet~\cite{isensee2021nnu}                  & 92.0    & 76.0       & 84.0   & 94.0   & 12.0      & 53.0 & 96.0    & 18.0     & 57.0    & 69.0 & 69.0 & 33.3 & 80.0  & 62.8 & 64.2                     \\
TransUNet~\cite{chen2021transunet,isensee2021nnu}              & \textbf{94.0}    & 78.0       & 86.0   & 94.0   & 22.0      & 58.0 & \textbf{96.0}    & 18.0     & 57.0    & 60.0 & 80.0 & 40.0 & 80.0  & 65.0 & 66.5                     \\ \hline
Ours                   & 90.0    & \textbf{84.0}       & \textbf{87.0}   & \textbf{94.0}   & \textbf{36.0}      & \textbf{65.0} & 82.0    & \textbf{48.0}     & \textbf{65.0}    & 60.0 & \textbf{80.0} & 40.0 & \textbf{86.7}  & 66.7 & \textbf{70.9}                     \\ 
\toprule
\end{tabular}
\label{tab:diag}
\vspace{-0.2cm}
\end{table*}

\begin{table}[]
\small
  \centering
    \caption{Ablation study on the representation of tumor queries. Average detection sensitivity (\%) and specificity (\%), and voxel-level tumor Dice scores are reported.  }
\tablestyle{2.4pt}{1.1}
\begin{tabular}{c|ccc}
\toprule
         & Sensitivity & Specificity & Dice \\ \hline
Plain    &    89.5   &   76.1    & 0.605 \\ 
Parallel &     90.1  &   78.9    &  0.608    \\
Hierarchy (Ours) &     93.3        &  81.7       &  0.628\\ \toprule
\end{tabular}\label{tab:ablation}
\vspace{-0.2cm}
\end{table}

\begin{table}[]

\small
  \centering
    \caption{Efficiency comparison. CancerUniT is 4.5x faster and 8x lighter than the assembly of single-tumor expert models (8-nnUNet).}
\tablestyle{2.4pt}{1.1}
\begin{tabular}{c|c|c}
\toprule
   Model               & Speed & Params \\ \hline
8-nnUNet ensemble & 187s    & 246.24M     \\ 
DeepLesion model~\cite{yan2018deep}        & 17s   & 70.94M \\ 
LENS (train on our data)~\cite{yan2021learning}             & 17s   & 70.94M \\ 
nnUNet~\cite{isensee2021nnu}             & 22s    & 30.78M     \\ 
TransUNet~\cite{chen2021transunet,isensee2021nnu}     & 25s    & 38.53M     \\ \hline
Ours              & 42s    & 30.87M     \\ 
\toprule
\end{tabular}\label{tab:effi}
\vspace{-0.4cm}
\end{table}

\textbf{Patient-level tumor detection per organ.} 
This task aims at the evaluation of whether the model can correctly localize and identify an existing tumor (agnostic of subtypes) or generate false positive tumor predictions in the normal controls. 
For example, if a patient has a tumor annotated in the liver, a true positive prediction means that the model predicts a liver tumor that overlaps (Dice $>$ 0) the ground-truth tumor annotation. We report the sensitivity for each organ and the specificity for normal controls in the test set. 

As shown in Table~\ref{tab1}, our model outperforms all the baseline models in terms of average sensitivity and specificity. Compared to the 8-nnUNet ensemble, our model has substantial improvement in the sensitivity of detecting colorectum (+18\%), lung (+20\%), and liver (+8\%) tumors, and the overall specificity (+21\%). We also observe improvements in these organs of other unified models, i.e., nnUNet and TransUNet, which demonstrate that the unified training of multi-organ multi-tumor segmentation will benefit almost every separate task, except for breast tumor (-2\%). Without seeing other organs and tumors, the separately trained models have many more false positives than unified models, with a much lower specificity of 59.2\%. 

Without seeing our data, the DeepLesion~\cite{yan2018deep} model has a moderate average sensitivity (64.0\%) and low specificity (4.2\%), hardly applicable to the real clinical scenario under such a high false positive rate. After training on our data with a leading lesion detection algorithm LENS~\cite{yan2021learning}, the sensitivity for colorectum, esophagus, and stomach, as well as the specificity are substantially improved; nevertheless, these are still lower than the segmentation-based models. These comparisons demonstrate that solving the tumor detection task as semantic segmentation is superior to using object detection methods. 

\textbf{Class-agnostic lesion-level tumor detection.} 
In lesion or tumor-level evaluation, we combine all lesions into one class and extract the lesion instances from the segmentation masks of ground-truth and predictions to compute the overall precision and recall. If a predicted lesion instance mask overlaps a ground-truth lesion, we count this prediction as true positive. As shown in Table~\ref{tab:lesion}, our approach has the highest precision and recall among all the methods. Both the DeepLesion models and the 8-nnUNet ensemble models have a large number of false positives, resulting in low precision. Similar to patient-level results, semantic segmentation algorithms generally do better than object detection methods. Our model outperforms the unified nnUNet model by approximately 5\% in precision and 4\% in recall.

{\bf Tumor segmentation.} This task focuses on the tumor segmentation quality, where our model still ranks as the top in segmentation Dice score per organ, as shown in Table~\ref{tab:seg_organ}. Here, we still ignore the subtype of the tumor and treat the tumors in the same organ with the same label. We only compare our model with the segmentation baselines, not the detection models (DeepLesion and LENS). Similar to tumor detection, the second best is the TransUNet model, and the unified nnUNet is better than its single counterpart. The improved performance of our model and TransUNet model illustrates that enhancing the CNN feature extraction with attentions will benefit multi-tumor segmentation. This observation is in line with our assumption that our query-based Transformer better explores the similarity between the inter-organ tumors, thus mutually improving the pixel-level texture differentiation of all tumors.

{\bf Tumor diagnosis.} Our third evaluation focuses on the diagnostic ability to differentiate different types of tumors on the four organs, i.e., pancreas, esophagus, stomach, and liver, where we have tumor subtypes including cancer and non-cancer. As shown in Table~\ref{tab:diag}, our method achieves the highest overall diagnosis performance of 70.9\%. Different from previous tumor detection and segmentation results, the single expert model is the second best (68.1\%), demonstrating its strong baseline performance on the diagnosis on single organs. The unified nnUNet model has a substantial performance drop (-4\%) compared to its separately trained counterpart. We hypothesize that this is due to the difficulty of multi-task training. With only voxel-wise supervision, a vanilla unified nnUNet is hard to well recognize numerous subtypes of tumors for accurate diagnosis. In contrast, our model is capable of exploiting the relationship between different tumor diagnosis tasks with our query hierarchy, thus maintaining high performance and even improving over the single expert models.

\begin{figure}[!tbp]
\centering
\includegraphics[width=\linewidth]{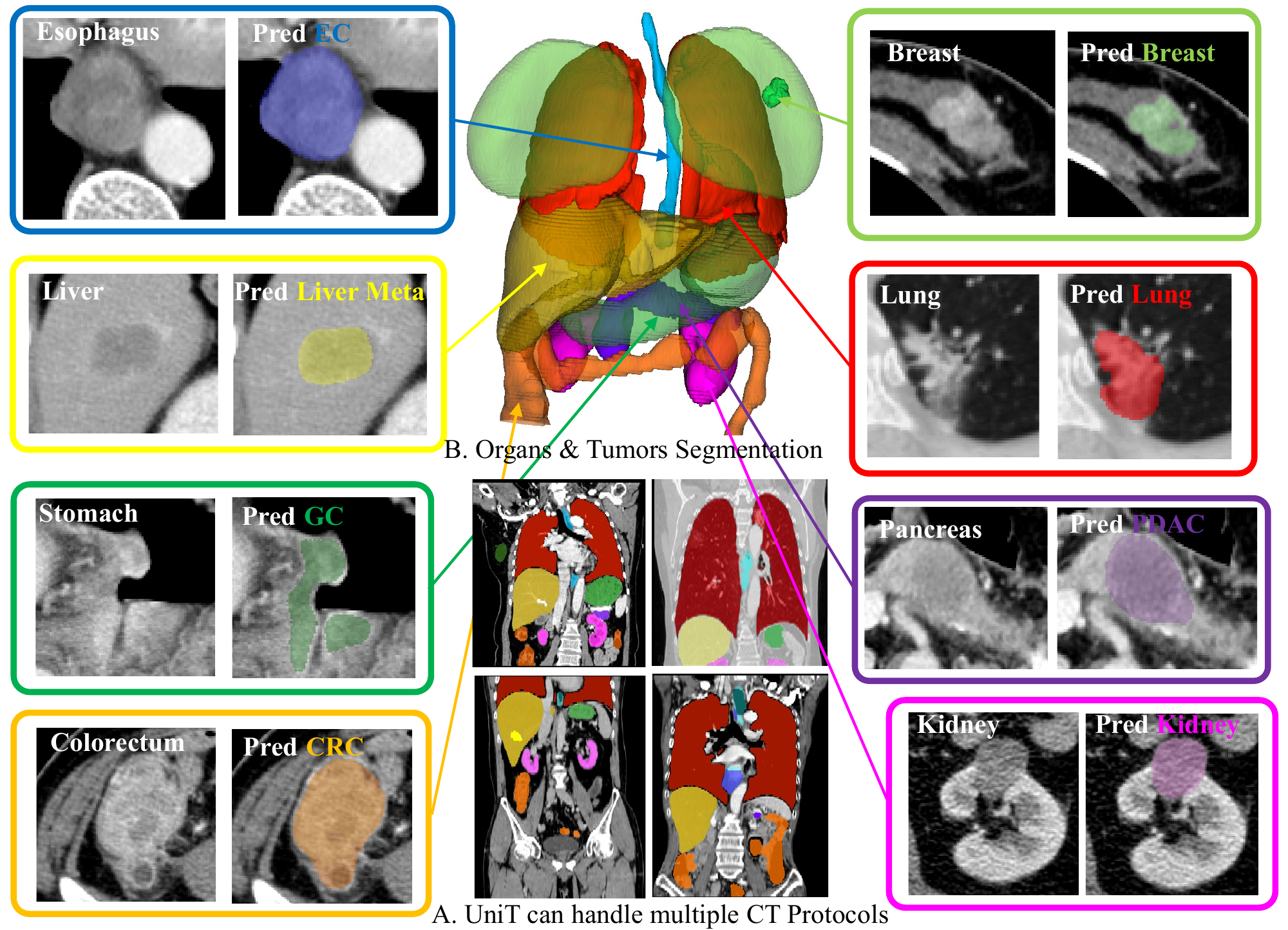}
\vspace{-0.4cm}
\caption{(A) Our model can handle multiple CT protocols representing the real-world clinical practice. (B) An example of the 3D masks of 8 organs and a breast tumor; and examples of 8 types of tumors being detected, segmented, and diagnosed by our CancerUniT. }
\label{fig:visual_examples}
\vspace{-0.2cm}
\end{figure}

\if
{\bf Universal Cancer Screening: CT \textit{vs} Blood Test.}
Blood test is now one of the most attractive tools for non-invasive multi-organ cancer screening~\cite{cohen2018detection_blood1,lennon2020feasibility,klein2021clinical_blood2}. CT scanning had been considered historically for the same task, but was limited by its insufficient sensitivity and specificity \cite{ahlquist2018universal}.
AI reading in CT as an alternative opportunistic screening tool, our approach also has strong clinical potential for cancer detection screening. The advantage of CT is that this protocol is already an indispensable diagnostic imaging for cancer, but a positive blood test result requires further examinations for confirmation. With our model, clinicians have direct visual analyses of the detected cancer sites and mis-detections of cancer can be largely reduced. No additional cost is needed under the opportunistic CT screening protocol whereas a single blood test can usually take $\sim1000$ US dollars. 

For relative performance comparison to CancerSeek~\cite{cohen2018detection_blood1}, i.e., cancer vs. normal, our method has higher sensitivity levels in detecting six out of seven types of cancers: approximately for stomach (+18\%), pancreas (+24\%), esophagus (+26\%), colorectum (+35\%), lung (+34\%), and breast (+57\%). Our averaged patient-level cancer detection sensitivity is 94\% versus 70\% in~\cite{cohen2018detection_blood1}. The test specificity for normal cases in venous CT is 100\% (blood test$>$99\%). We acknowledge that the results of the representative blood test~\cite{cohen2018detection_blood1} and ours may not directly comparable since different test data are used. Nevertheless, the rough comparison shows the high accuracy of CT+AI solution, and thus may re-open doors for multi-cancer screening by CT \cite{ahlquist2018universal}.  
\fi

{\bf Ablation study.} We perform the ablation study on the representation of tumors (Table \ref{tab:ablation}). We compare the other two representations: (1) parallel representation: the detection queries and diagnosis queries are organized as two groups in parallel, without structural connection. (2) plain representation: only diagnosis queries are used in our framework, while the prediction for a major tumor $\mbf{m}_i$ in the detection branch is directly obtained by merging several subtypes of tumors $\mbf{n_i}$ in the diagnosis branch.

{\bf Efficiency.} We compare the efficiency among various models in both inference speed and model size (number of parameters) as illustrated in Table \ref{tab:effi}.  CancerUniT is 4.5x faster and 8x lighter than the assembly of single-tumor expert models.

Visual results in Fig.~\ref{fig:visual_examples} shows that our model can handle multiple CT protocol, and is capable to detect, segment, and diagnose 8 types of major cancers. {\bf Generalizability} to public dataset is shown in Supplementary.

\section{Conclusion}

In this paper, we propose a single unified tumor Transformer (CancerUniT) model to detect, segment and diagnose eight common cancers using 3D CT scans, for the first time. CancerUniT is a query-based Transformer and offers a novel clinically inspired hierarchical tumor representation, with a dual-task query decoding stage for segmentation mask generation. We curate a large collection of CT scans with high clinical quality from 10,673 patients, including eight major types of cancers and occurring non-cancer tumors (pathology-confirmed and manually annotated). Extensive quantitative evaluations have demonstrated the promising performance of our new model. This moves one step closer to a universal high performance cancer screening AI tool.



\smallskip\noindent\textbf{Acknowledgments.}
Jieneng Chen and Alan Yuille in this project were partially funded by a 2023 Patrick J. McGovern Foundation award.

{\small
\bibliographystyle{ieee_fullname}
\bibliography{egbib}
}

\clearpage

\section*{Appendix: CancerUniT}
\appendix

\noindent\textbf{Abstract.} This document contains the Supplementary Materials for the ICCV 2023 paper "CancerUniT: Towards a Single Unified Model for Effective Detection, Segmentation, and Diagnosis of Eight Major Cancers Using a Large Collection of CT Scans". It covers the model generalizability to public dataset, (\S\ref{sec:gene}), model instantiation details (\S\ref{sec:inst}), semantic segmentation results of full spectrum tumors (\S\ref{sec:segfull}), and the qualitative results (\S\ref{sec:viz}).

\section{Generalizability to Public Dataset}
\label{sec:gene}
Our method aims at holistically modeling the multiple cancer screening problem versus non-cancer. However, to the best of our knowledge, no public dataset is suitable for such problems. Nevertheless, our trained model generalizes well on three public single-tumor datasets including MSD pancreas, liver and lung dataset, as shown in  Table.~\ref{tab:gel}. It is worth noting that our model inference directly without extra training, whereas the 3 single-nnUNets is trained on the MSD dataset with domain knowledge. To be specific, the experiment of 3 single-nnUNet is conducted with 5-fold cross-validation, and our CancerUniT is tested on the same validation set. 

Despite not having any prior knowledge of the data distribution, our proposed UniT model effectively suppresses the single-tumor expert model, achieving an average tumor detection sensitivity improvement of 3.1\%. Our results demonstrate the efficacy of our proposed method for addressing the tumor detection problem without the need for a specific dataset. The ability to generalize well on public datasets and suppress the single-tumor expert model underscores the potential of our approach to be used as a practical solution for universal cancer screening and diagnosis.

\begin{table}[h!]
\label{tab:gel}
\small
  \centering
    \caption{Generalizability to 3 Public MSD dataset~\cite{antonelli2022medical}. Average detection sensitivity is reported. Our model inference directly, whereas 3 single-nnUNets are trained on the MSD dataset.}
    \vspace{0.2cm}
\tablestyle{2.4pt}{1.1}
\begin{tabular}{c|c|c|c|c|c|c}
\toprule
                                                                   & \begin{tabular}[c]{@{}c@{}}Pancreatic \\ tumor\end{tabular} & \begin{tabular}[c]{@{}c@{}}Liver \\ tumor\end{tabular} & \begin{tabular}[c]{@{}c@{}}Lung \\ tumor\end{tabular} & Avg    & Speed & Param   \\ \hline
\begin{tabular}[c]{@{}c@{}}single-nnUNet \\ (trained)\end{tabular} & 88\%                                                        & 97\%                                                   & 90.5\%                                                  & 91.8\% & 66s   & 92.34M \\ \hline
\begin{tabular}[c]{@{}c@{}}Ours\\ (test)\end{tabular}              & 94.7\%                                                        & 93.1\%                                                   & 97\%                                                  & 94.9\% & 42s   & 30.87M  \\ 
 \toprule
\end{tabular}
\end{table}

\section{Model Instantiation Details}
\label{sec:inst}
In our CancerUniT, the hidden dimension of query is set to 32, such that the detection query $\mathbf{A}^j \in \mathbb{R}^{4 \times 32}$, the diagnosis query $\mathbf{B}^j \in \mathbb{R}^{10 \times 32}$, the shared query $\mathbf{S}^j \in \mathbb{R}^{12 \times 32}$. We adopt nnUNet~\cite{isensee2021nnu} as the backbone to extract multi-scale features $\mathbf{F}=[\mathbf{F}^1, \mathbf{F}^2, \mathbf{F}^3, \mathbf{F}^4]$. Note, $\mathbf{F}^j \in \mathbb{R}^{d \times (D\times H \times W)}$ is flatten and projected from intermediate spatial feature $\mathbf{\hat{F}}^j \in \mathbb{R}^{C \times D\times H \times W}$. In specific, $\mathbf{F}^1 \in \mathbb{R}^{32 \times (48\times 192 \times 192)}$, $\mathbf{F}^2 \in \mathbb{R}^{32 \times (48\times 96 \times 96)}$, $\mathbf{F}^3 \in \mathbb{R}^{32 \times (24\times 48 \times 48)}$, and $\mathbf{F}^4 \in \mathbb{R}^{32 \times (12\times 24 \times 24)}$. The total number of Transformer layer is set to 3, each of which contains a multi-head cross-attention, a multi-head self-attention, and a feed-forward network. Note, in the inference stage, the tumor segmentation maps are extract to generate the tumor instances with class labels, where those tumor instances with less than 200 voxels are discarded.

\begin{table*}[ht]
\small
  \centering
    \caption{Voxel-level semantic segmentation results of full spectrum tumors. The Dice coefficient is reported. Note: the Dice values are calculated in a semantic manner, e.g., the HCC voxel is correctly segmented as the HCC subtype (not other liver tumor subtypes or other tumor types) by the semantic segmentation. }
    
\tablestyle{2.3pt}{1.0}
\begin{tabular}{c|ccc|ccc|ccc|ccccc|c}
\toprule
\multirow{2}{*}{Model} & \multicolumn{3}{c|}{Pancreas}              & \multicolumn{3}{c|}{Eso}                      & \multicolumn{3}{c|}{Stomach}             & \multicolumn{5}{c|}{Liver}                               & \multicolumn{1}{c}{\multirow{2}{*}{Average}} \\
                       & PDAC  & nonPDAC & \multicolumn{1}{c|}{avg} & EC & nonEC & \multicolumn{1}{c|}{avg} & GC    & nonGC & \multicolumn{1}{c|}{avg} & HCC   & ICC   & Meta  & Heman & \multicolumn{1}{c|}{avg} & \multicolumn{1}{c}{}                         \\ \hline
8-nnUNet ensemble      & 0.750 & 0.525   & 0.638                    & 0.770  & 0.433     & 0.602                    & 0.441 & 0.099  & 0.270                    & 0.489 & 0.552 & 0.296 & 0.784 & 0.530                    & 0.510                                        \\
nnUNet~\cite{isensee2021nnu}                 & 0.758 & 0.534   & 0.646                    & 0.739  & 0.207     & 0.473                    & 0.453 & 0.068  & 0.261                    & 0.410 & 0.481 & 0.306 & 0.739 & 0.484                    & 0.466                                        \\
TransUNet~\cite{chen2021transunet}              & 0.749 & 0.553   & 0.651                    & 0.744  & 0.321     & 0.533                    & 0.473 & 0.128 & 0.301                    & 0.411 & 0.503 & 0.353 & 0.717 & 0.496                    & 0.495                                        \\ \hline
Ours                   & 0.728 & 0.560   & 0.644                    & 0.738  & 0.457     & 0.597                    & 0.389 & 0.187 & 0.288                    & 0.368 & 0.666 & 0.305 & 0.773 & 0.528                    & 0.514                                        \\ 
\toprule
\end{tabular}
\label{tab:segfull}
\end{table*}

\section{Semantic Segmentation Results of Full Spectrum Tumors}
\label{sec:segfull}

We conducted an evaluation of our model's performance on the semantic segmentation of full spectrum tumors, which is a challenging task that involves the segmentation of multiple tumor subtypes within an organ. The quality of the multi-class tumor segmentation was assessed using the multi-class Dice score, where each subtype of the tumor was treated as an independent semantic class.

Our model outperformed the segmentation baselines and achieved the highest average segmentation Dice score, as demonstrated in Table~\ref{tab:segfull}. Notably, our model was not compared with detection models such as DeepLesion and LENS, as these models are not designed for semantic segmentation tasks.

Our findings suggest that enhancing the query hierarchy in our model can improve the semantic segmentation of full spectrum tumors. This observation is in line with our assumption that our query-based Transformer model can more effectively explore the similarity between intra-organ tumor subtypes, leading to improved segmentation performance. Overall, our evaluation provides evidence that our proposed model can effectively address the challenges of multi-class tumor segmentation in the context of full spectrum tumors.

\section{Universal Cancer Screening: CT \textit{vs} Blood Test.}
Blood test is now one of the most attractive tools for non-invasive multi-organ cancer screening~\cite{cohen2018detection_blood1,lennon2020feasibility,klein2021clinical_blood2}. CT scanning had been considered historically for the same task, but was limited by its insufficient sensitivity and specificity \cite{ahlquist2018universal}.
AI reading in CT as an alternative opportunistic screening tool, our approach also has strong clinical potential for cancer detection screening. The advantage of CT is that this protocol is already an indispensable diagnostic imaging for cancer, but a positive blood test result requires further examinations for confirmation. With our model, clinicians have direct visual analyses of the detected cancer sites and mis-detections of cancer can be largely reduced. No additional cost is needed under the opportunistic CT screening protocol whereas a single blood test can usually take $\sim1000$ US dollars. 

For relative performance comparison to CancerSeek~\cite{cohen2018detection_blood1}, i.e., cancer vs. normal, our method has higher sensitivity levels in detecting six out of seven types of cancers: approximately for stomach (+18\%), pancreas (+24\%), esophagus (+26\%), colorectum (+35\%), lung (+34\%), and breast (+57\%). Our averaged patient-level cancer detection sensitivity is 94\% versus 70\% in~\cite{cohen2018detection_blood1}. The test specificity for normal cases in venous CT is 100\% (blood test$>$99\%). We acknowledge that the results of the representative blood test~\cite{cohen2018detection_blood1} and ours may not directly comparable since different test data are used. Nevertheless, the rough comparison shows the high accuracy of CT+AI solution, and thus may re-open doors for multi-cancer screening by CT \cite{ahlquist2018universal}.

\section{Qualitative Results}
\label{sec:viz}

\begin{figure*}[t!]
    \small
    \centering
    \includegraphics[width=0.8\textwidth]{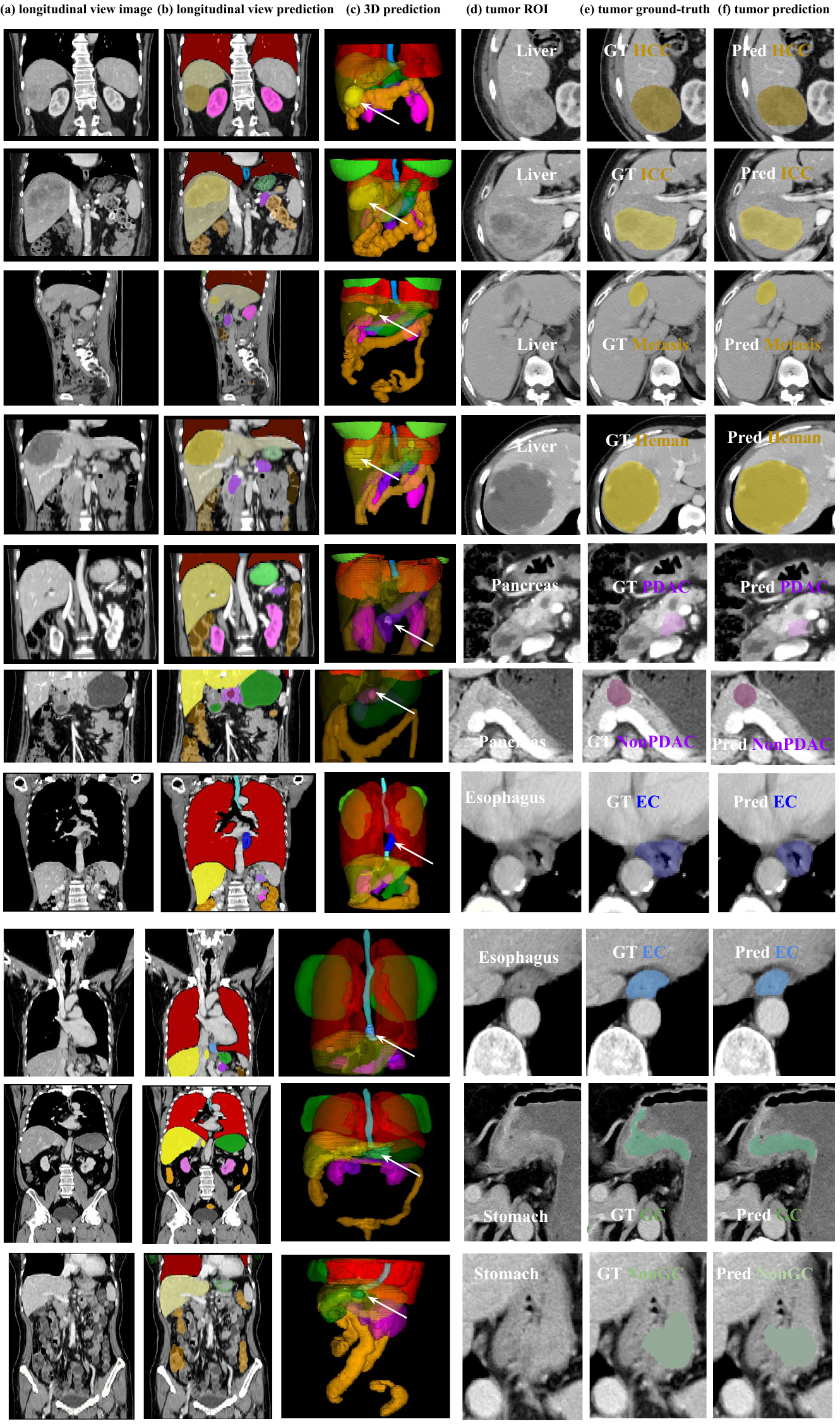}
    \caption{Qualitative results of full spectrum tumors in the test set being segmented and diagnosed by our method (best viewed in color). }
    \label{fig:supp_vis}
\end{figure*}

We provide more qualitative results of full spectrum tumors in the test set being segmented and diagnosed by our method as shown in Fig.~\ref{fig:supp_vis}. The results demonstrate that our method can not only segment the tumor region well but also predict the class of tumor subtype correctly.

\end{document}